\documentclass[twocolumn,english,aps,prb,floatfix]{revtex4}
\usepackage[T1]{fontenc}
\usepackage[latin1]{inputenc}
\usepackage{amsmath}
\usepackage{graphicx}
\usepackage{amssymb}

\pdfoutput=1

\makeatletter

\usepackage{babel}
\makeatother
\begin{document}

\title{Spectroscopy, upconversion dynamics, and applications of Er$^{3+}$-doped
low-phonon materials}

\author{Angel J. Garcia-Adeva}

\affiliation{Departamento de Fisica Aplicada I, E.T.S. Ingenieria de Bilbao, Alda.
Urquijo s/n, 48013 Bilbao, Spain}

\email{angel.garcia-adeva@ehu.es}

\begin{abstract}
In this work I summarize some of the recent work carried out by our
group on the upconversion dynamics of Er$^{3+}$-doped potassium lead
halide crystals, which possess very small phonons and present very
efficient blue and green upconversion. Furthermore, a non-conventional
application of these RE-doped low-phonon materials in optical refrigeration
of luminescent solids is also discussed, paying especial attention
to new pathways for optical cooling that include infrared-to-visible
upconversion. Finally, I conclude with some hints of what I think
it is the next step into improving the luminescence efficiency of
solids: the use of RE-doped nanoscale photonic heterostructures for
controlling the density of photonic states.
\end{abstract}
\maketitle

\section{Introduction}

Rare-earth-doped materials capable of converting infrared radiation
into visible light in an efficient way are potential candidates for
photonic applications in several areas such as color displays, sensors,
detection of infrared radiation, and upconversion lasers \cite{key-1}.
Indeed, upconversion luminescence is considered as a promising solution
to obtain efficient visible lasers pumped by commercial infrared laser
diodes. The main shortcoming for the development of such applications
is the need of host materials with low-phonon energy that leads to
a significant reduction of the multiphonon relaxation rates and, thus,
allows for an increased lifetime of some excited levels that can relax
radiatively or can store energy for further upconversion, cross-relaxation,
or energy-transfer processes.

Efficient ir-to-vis upconversion and laser action has been demonstrated
in chlorides, bromides, and iodides compounds doped with different
RE ions \cite{key-7,key-8,key-9,key-10,key-11,key-12}. In this low-phonon-energy
materials, due to very small multiphonon relaxation rates and the
long lifetime of the $^4$I$_{9/2}$ level, this one acts as an intermediate
state for upconversion, leading to a number upconversion mechanisms.
For instance, blue, green, and red emissions from the $^2$H$_{9/2}$
level by sequential upconversion excitation from the $^4$I$_{9/2}$
level has been demonstrated. Unfortunately, one drawback of chloride
and bromide systems is that these materials usually present poor mechanical
properties, moisture sensitivity, and are difficult to synthesize.
In this regard, an important advance in the search for new low-phonon-energy
materials has been the identification of potassium lead halide crystals
KPb$_{2}X_{5}$ ($X$=Cl, Br) as new low-energy phonon hosts for RE
ions \cite{key-12,key-13,key-14,key-15,key-16,key-17,key-18,key-19}.
These crystals are non-hygroscopic and readily incorporate RE. The
maximum phonon energy are 203 cm$^{-1}$ and 138 cm$^{-1}$ for the
KPb$_{2}$Cl$_{5}$ and KPb$_{2}$Br$_{5}$ crystals, respectively.
These small phonons are the main reason why efficient ir-to-vis upconversion
in Pr$^{3+}$-doped and Yb$^{3+}$-Pr$^{3+}$-codoped KPb$_{2}$Cl$_{5}$,
in Er$^{3+}$-doped KPb$_{2}$Cl$_{5}$, Nd$^{3+}$-doped KPb$_{2}$Cl$_{5}$,
and Er$^{3+}$-doped KPb$_{2}$Br$_{5}$ has been recently found by
our group and, also, why we have been able to use one of these matrices
for laser cooling of luminescent solids.

The structure of this paper is as follows: in the next section, I
will briefly summarize the main features of the ir-to-vis upconversion
processes found in these systems. Then, I will explain how these materials
could be used for the development of optical refrigeration applications
of luminescent solids. I will end up this work explaining how I think
one can control, from a general perspective, the luminescence efficiency
of these solids and, in particular, the cooling efficiency, by using
photonic nanostructures doped with optically active ions (such as
RE) as a substrate for these applications.

\section{Upconversion dynamics of RE-doped potassium lead halide crystals}

Single crystals of non-hygroscopic Er$^{3+}$-doped ternary potassium
lead chloride (KPC) and bromide (KPB) grown by the Bridgman technique
were investigated. The maximum Er$^{3+}$ concentration of the samples
was about 1.0\%. Conventional absorption spectra between 175 and 3000
nm were performed with a Cary 5 spectrophotometer. The steady-state
emission measurements were made with a Ti-sapphire ring laser 0.4
cm$^{-1}$ linewidth as exciting light source. The excitation beam
was focused on the crystal by using a 50 mm focal lens. The fluorescence
was analyzed with a 0.25 m monochromator, and the signal was detected
by a Hamamatsu R928 photomultiplier and finally amplified by a standard
lock-in technique. Lifetime measurements were obtained by exciting
the sample with a dye laser pumped by a pulsed nitrogen laser and
a Ti-sapphire laser pumped by a pulsed frequency-doubled neodymium-doped
yttrium aluminum garnet Nd:YAG laser 9 ns pulse width, and detecting
the emission with Hamamatsu R928 and R5509-72 photomultipliers. The
data were processed by a Tektronix oscilloscope.

\begin{figure}
\begin{centering}\includegraphics{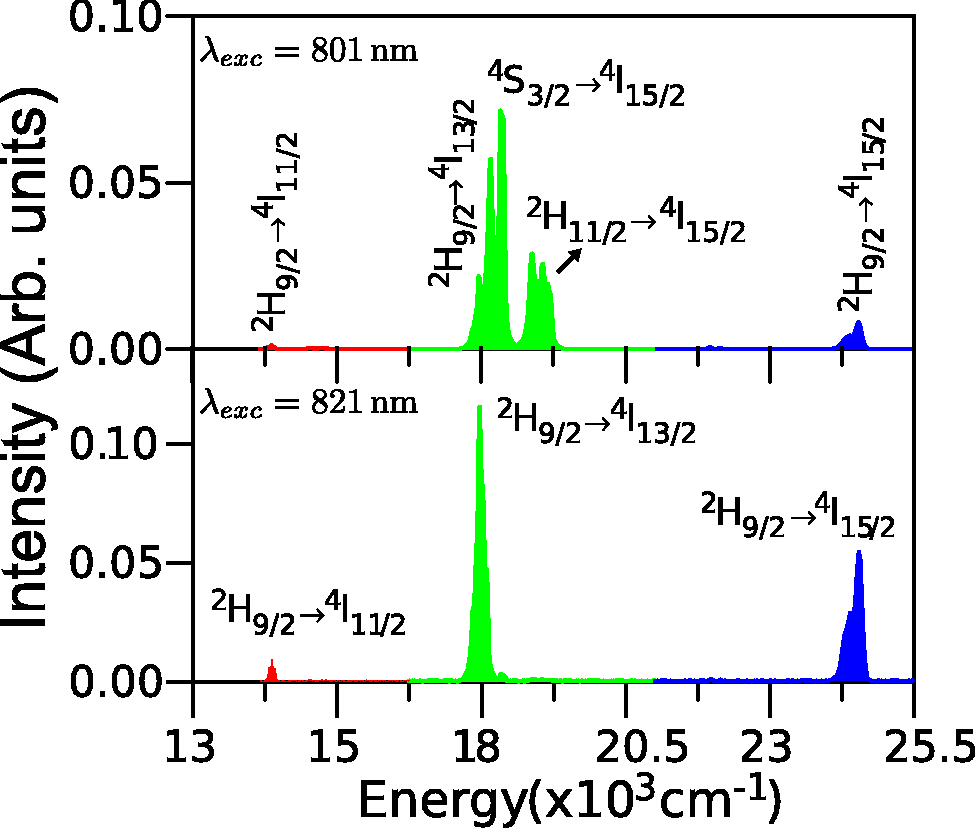}\end{centering}
\caption{\label{fig:kpcupconversion}Er$^{3+}$-doped KPC upconverted emission
spectra upon excitation at 801 nm (upper panel) and 821 nm (lower
panel). The Er concentration was 0.5\%.}
\end{figure}

\begin{figure}
\begin{centering}\includegraphics{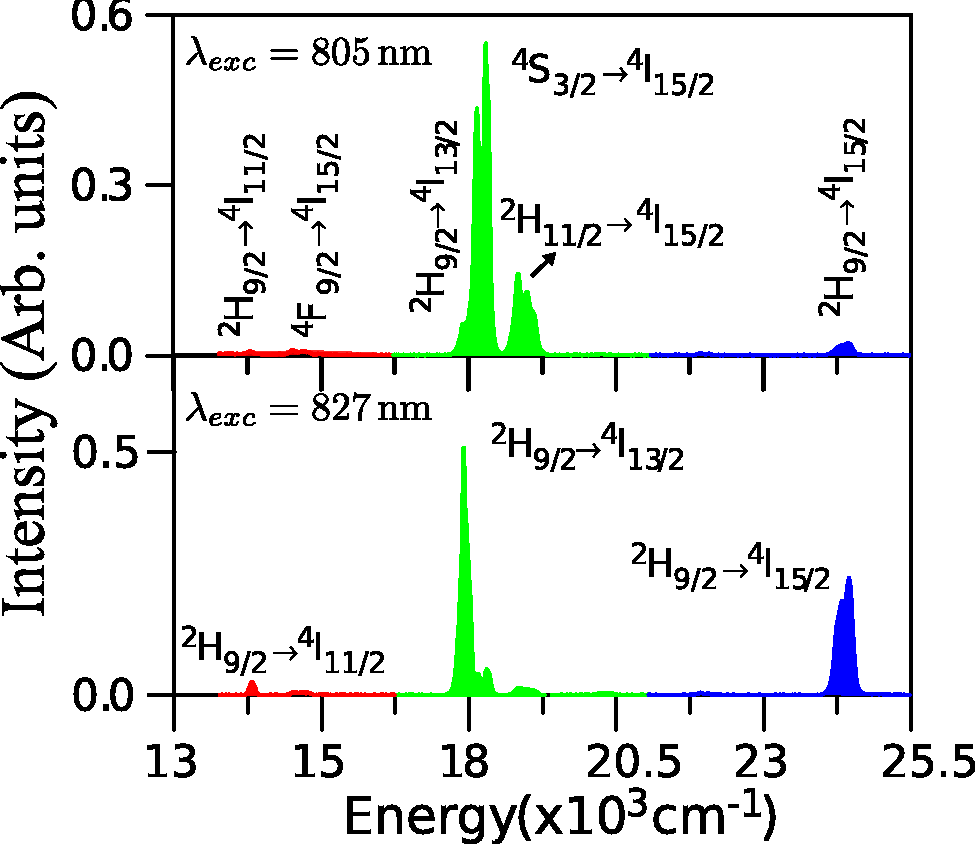}\end{centering}
\caption{\label{fig:kpbupconversion}Er$^{3+}$-doped KPB upconverted emission
spectra upon excitation at 805 nm (upper panel) and 827 nm (lower
panel). The Er concentration was around 1.0\%.}
\end{figure}

It is experimentally found that efficient red, green, and blue upconverted
emissions are observed when matrices of KPC and KPB doped with Er$^{3+}$
--up to a 1.0 \% concentration-- are pumped with cw laser radiation
of an energy that corresponds to the energy difference between the
$^4$I$_{9/2}$ manifold and the $^4$I$_{15/2}$ ground-state manifold.
These results are summarized in figures \ref{fig:kpcupconversion}
and \ref{fig:kpbupconversion} for the KPC and KPB systems, respectively.
In particular, for the Er$^{3+}$-doped KPC when we pump resonantly
with the barycenter of the $^4$I$_{9/2}$ level (801 nm), there are
red, green, and blue upconverted emissions coming out from the $^2$H$_{9/2}$
and $^4$S$_{3/2}$ levels, as shown in the upper panel of figure \ref{fig:kpcupconversion}.
On the other hand, when we pump just below the $^4$I$_{9/2}$ level,
we only get upconverted red, green, and blue emissions from the $^2$H$_{9/2}$
manifold, as can be seen in the lower panel of figure \ref{fig:kpcupconversion}.%

The situation is very similar in the case of the KPB compound (see
fig.~\ref{fig:kpbupconversion}). The only difference is that there
is also a small amount of red upconverted emission coming from the
$^4$F$_{9/2}$ level when we excite resonantly into the $^4$I$_{9/2}$
level (upper panel in figure \ref{fig:kpbupconversion}). The upconverted
emission spectrum for non-resonant excitation just below this band,
however, is totally analogous to the KPC case.

Apart from assessing the capability of these systems for efficient
ir-to-vis upconversion, we were also very interested in elucidating
the different mechanisms responsible for this upconversion. In this
regard, a first clue to identify these is provided by the study of
the power dependence of the upconverted emission intensity. The result
of such an study shows that this quantity follows a power law dependence
which is closer to a quadratic law rather than a linear one \cite{key-12,key-21}.
This is a clear indication that two photons are involved in the upconversion
process. Another useful tool to investigate this is the study the
upconverted population dynamics after pulsed excitation. This technique
allows one to easily distinguish between an excited state absorption
(ESA) process and an energy transfer upconversion (ETU) process \cite{key-20}.
In the first case, one electron in an excited state (the $^4$I$_{9/2}$
level, for example) absorbs a pumping photon and is promoted to a
higher excited state (the $^2$H$_{9/2}$, for example). Later, when
this electron decays spontaneously, it emits a photon of frequency
twice the frequency of the pumping photons. In the second case, an
electron in one ion initially in an excited state, decays spontaneously
by exchanging a photon with a second electron in the same excited
state of another ion and, thus, this later one is promoted to a higher
excited state. Later, when this electron decays radiatively, it emits
a photon of frequency twice the frequency of the incident photons.
As said above, the population dynamics after pulsed excitation in
these two cases is quite different. The typical signature of an ESA
process is an exponential decay, whereas the signature of an ETU process
is a rise followed by a an exponential decay.%
\begin{figure*}
\begin{centering}\includegraphics{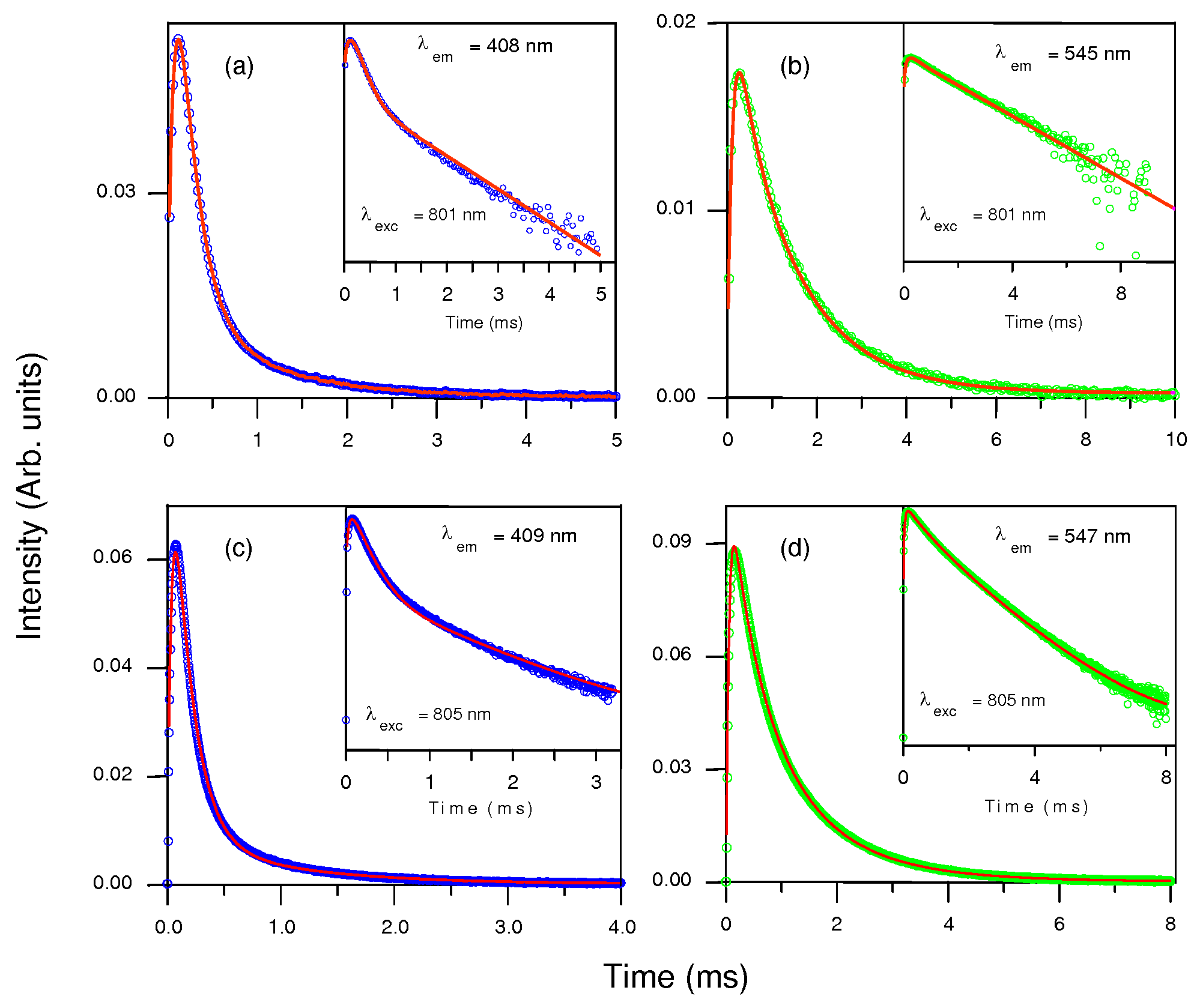}\end{centering}
\caption{\label{fig:kpxfit}(a) Blue and (b) green upconverted emission intensity
as a function of time for non-resonant excitation into the $^4$I$_{9/2}$
band together with the fit (solid lines) to our theoretical model
for Er$^{3+}$-doped KPC. (c) and (d): analogous to (a) and (b), respectively,
for Er$^{3+}$-doped KPB.}
\end{figure*}

Experimentally, we find that, in the particular case of Er$^{3+}$-doped
KPC, upon resonant excitation into the $^4$I$_{9/2}$ level (801 nm),
we get a typical ETU behavior for both the blue and green upconverted
emissions (see fig.~\ref{fig:kpxfit}a and \ref{fig:kpxfit}b). The
blue upconverted population initially increases and then it decays
following a two-exponential behavior. The green upconverted population
initially rises too and, then, it decays according to a single-exponential
behavior. Even though I will not delve into these details here, it
is important to stress that the rise and decay time constants are
sub-multiples of the lifetimes of the levels involved in the upconversion
processes. I refer the interested reader to references \onlinecite{key-12}
and \onlinecite{key-21} for more details. One the other hand, the
situation is radically different when we pump just below the $^4$I$_{9/2}$
level using a pumping frequency such that two pumping photons are
in resonance with the $^2$H$_{9/2}$ level. Under this condition,
the upconverted blue and green emissions show a clear ESA behavior
(not shown here, see refs.~\onlinecite{key-12} and \onlinecite{key-21}
for details). The results for the Er$^{3+}$-doped KPB system are
totally similar to the ones obtained for the KPC and can be seen in
figure \ref{fig:kpxfit}c and \ref{fig:kpxfit}d.

In order to theoretically investigate the mechanisms responsible for
the upconversion under non-resonant pumping, we performed a very simple
rate equation analysis for both the KPC and KPB systems. The details
of such approach can be found in refs.~\onlinecite{key-12} and \onlinecite{key-21}
and will be not reproduced here. For the present work, it suffices
to say that the analytical expressions obtained in this way are in
excellent agreement with the experimental data of the upconverted
population dynamics, as can be seen in figures \ref{fig:kpxfit}a
and \ref{fig:kpxfit}a and in figures \ref{fig:kpxfit}c and \ref{fig:kpxfit}d
for both the blue and green upconverted emissions in Er$^{3+}$-doped
KPC and KPB, respectively. This, in turn, is a strong support for
the validity of the proposed upconversion models.

\section{Laser cooling of luminescent solids}

\begin{figure}
\begin{centering}\includegraphics{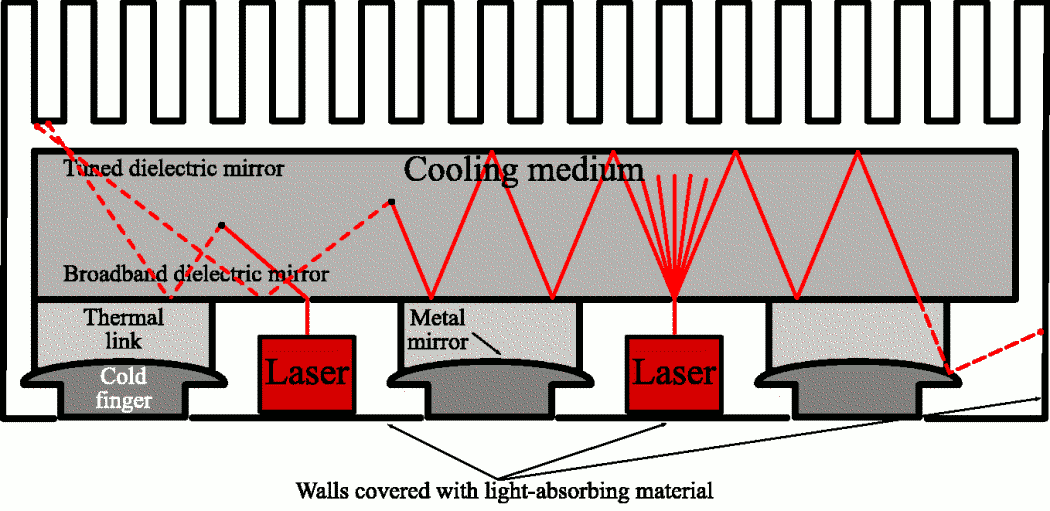}\end{centering}
\caption{\label{fig:cryocooler}A proposed design for an optical cryocooler
(based on a patent by Edwards et al., ref.~\onlinecite{key-28}).}
\end{figure}

In this section I will present a non-conventional application of the
Er$^{3+}$-doped low-phonon materials described in the previous section
for the development of optical cryocoolers. The physics behind laser
cooling of luminescent solids is based on what is known as an anti-Stokes
transition \cite{key-22}: an optically active ion absorbs a photon
from a pump laser beam and this excites an electron from the ground
state manifold to the excited-state one. In the process, one or more
phonons from the lattice are annihilated in order to make up for the
energy mismatch between the pump photon energy and the excited state
one. Later, when the electron decays spontaneously, it emits a fluorescence
photon of frequency slightly larger than the pump one, so that in
each of these events, a small amount of thermal vibrational energy
is removed from the system \cite{key-23,key-24,key-25}. Therefore,
if this process occurs in a cyclic way, we have an optical refrigerator
with a maximum theoretical cooling efficiency given by \cite{key-24}\begin{equation}
\eta_{\text{cool}}=\frac{\nu_{f}-\nu_{p}}{\nu_{p}}.\end{equation}
Of course, this effect could have many applications \cite{key-26}.
The simplest and probably most profitable one is for developing cryocoolers
for the microprocessors of personal computers. Indeed, current generation
CPU coolers have many mechanical parts and, thus, are subject to various
shortcomings: they generate a lot of vibrations, they are noisy, and
consume a lot of power. On the other hand, an optical cryocooler could
look like the device shown in figure \ref{fig:cryocooler} (see refs.~\onlinecite{key-27}
and \onlinecite{key-28}). It consists of an active medium that is
pumped by a diode laser and which is linked to a cold finger that
actually cools the CPU. The external walls of the device are covered
with a special coating that absorbs any fluorescence escaping from
the active medium. It is easy to realize that such a device would
have many advantages: it has no mechanical parts and, thus, there
are no vibrations, no noise, and --with the advent of the new generation
diode lasers-- it would consume very little power. Other important
applications of this type of laser cooling are, for example, the development
of radiation-balanced laser that use dual wavelength pumping to offset
the heat generated by the pump laser. Also, this could have many applications
in bioimaging and phototherapy, where this dual wavelength pumping
could also partially offset the heat that could otherwise damage the
living specimen under study.%

\begin{figure}
\begin{centering}\includegraphics{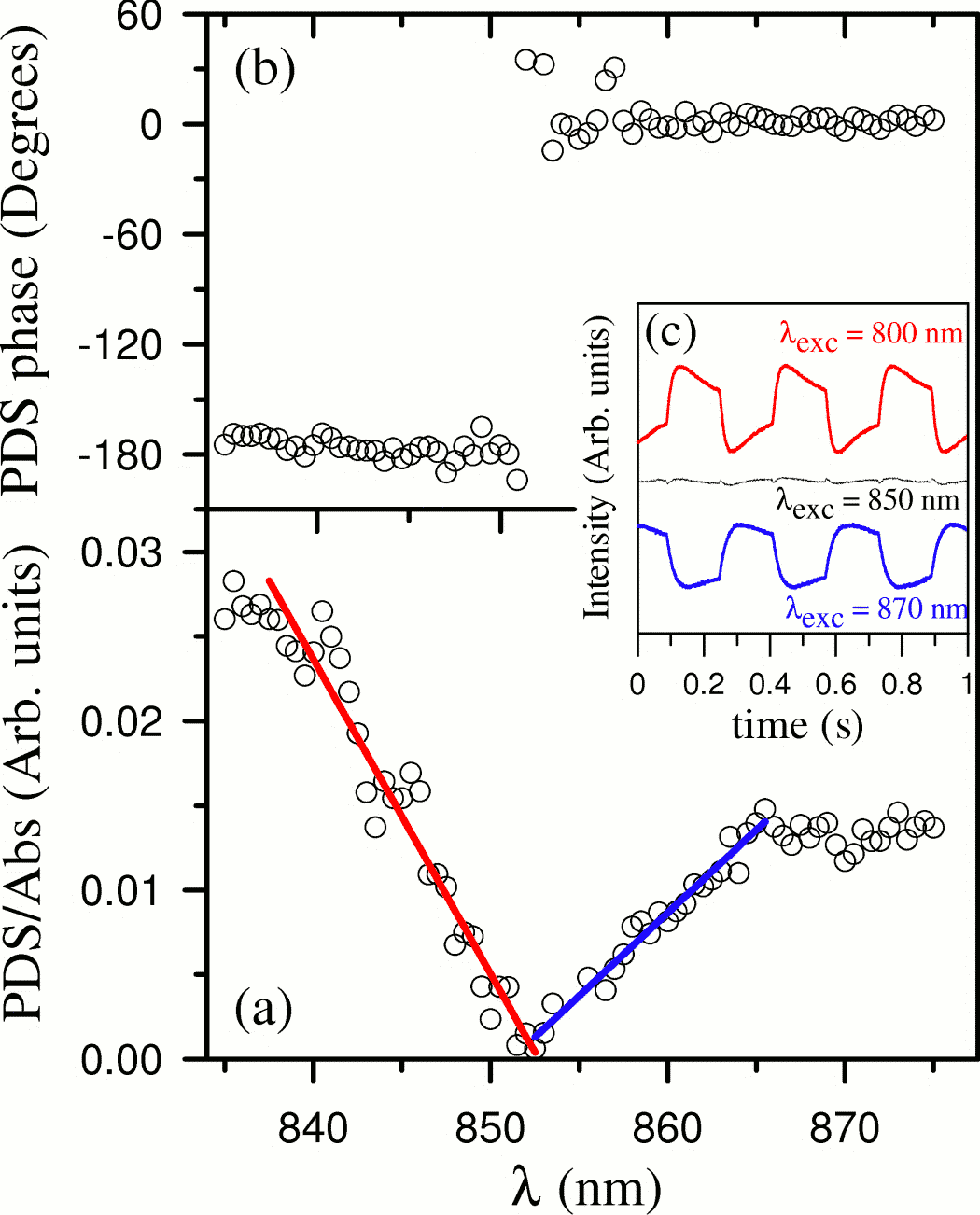}\end{centering}
\caption{\label{fig:pds}(a)Photothermal signal deflection amplitude normalized
by the sample absorption as a function of pumping wavelength for the
Er$^{3+}$:KPb$_{2}$Cl$_{5}$ crystal. (b) Phase of the photothermal
deflection signal as a function of pumping wavelength. (c) Photothermal
deflection signal waveforms in the heating (800 nm) and cooling (870
nm) regions and around the cooling threshold (850 nm).}
\end{figure}

\begin{figure}
\begin{centering}\includegraphics{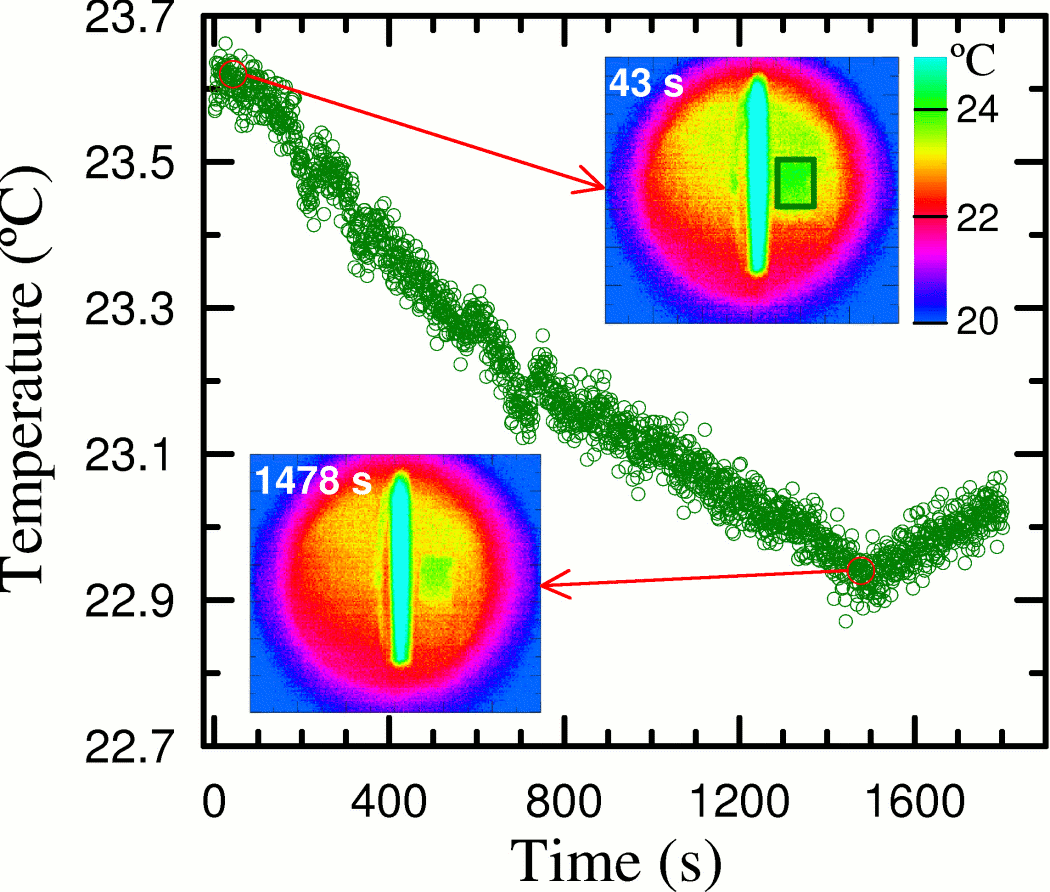}\end{centering}
\caption{\label{fig:kpcir}Time evolution of the average temperature of the
Er$^{3+}$:KPb$_{2}$Cl$_{5}$ at 870 nm. The insets show colormaps
of the temperature field of the whole system (sample plus cryostat)
at two different times as measured with the thermal camera. The rectangle
in the upper inset delimits the area used for calculating the average
temperature of the sample.}
\end{figure}

Unfortunately, these applications are still way ahead down the road
--even though there are some prototypes of optical cryocoolers made
at Los Alamos by Epstein and coworkers \cite{key-27}-- so, in the
meantime, a number of research groups are trying to investigate novel
materials amenable of efficient laser cooling \cite{key-23,key-24,key-29,key-30}.
There are two typical probes used in the lab to investigate this phenomenon.
On the one hand, in order to probe local or internal cooling, the
technique known as collinear photothermal deflection spectroscopy
is commonly used. This technique is based in the mirage effect, where
a probe laser beam probes the thermal lens (an index of refraction
gradient) created by a pump laser beam. When the pump laser is adequately
tuned so that there is a transition from heating to cooling, there
is a change in the direction in which the probe beam deviates. A typical
result obtained with the PDS technique can be seen in figure \ref{fig:pds}
for an Er$^{3+}$-doped KPC crystal with 0.5\% Er when pumping into
the $^4$I$_{9/2}$ band \cite{key-26}. Above the barycenter of this
band there is the Stokes region, in which heating occurs. The barycenter
appears in that figure as the frequency at which the PDS signal is
almost zero. Below this frequency, we enter in the anti-Stokes region,
in which cooling occurs. This shows up in the upper panel of that
figure as a 180º change in the phase of the PDS signal that clearly
indicates the transition from heating to cooling. Also, the transition
from heating to cooling can be noticed as the change in sign of the
waveforms captured with an oscilloscope, as shown in the (c) panel
of figure \ref{fig:pds}. Furthermore, from the slopes of the PDS
signal, one can estimate the cooling efficiency. For this particular
system, the cooling efficiency is very small and around 0.4\%. On
the other hand, in order to probe the presence of macroscopic or bulk
cooling --more useful, in fact, for practical applications-- commonly
an IR thermal camera is used. The typical pictures taken by this device
look like the ones shown in the insets of figure \ref{fig:kpcir}
for the same sample as before. The figure on the right inset corresponds
to an IR picture taken 43 seconds after the laser irradiation of the
sample started. The figure on the left inset corresponds to a picture
taken after 1483 seconds had passed since laser irradiation started.
It can be seen that the color of sample changes between these two
instants of time which, according to the color scale, indicates a
slight decrease of the temperature of the sample. Actually, it is
quite difficult to extract quantitative information from those pictures
directly and, therefore, it is better to calculate the average temperature
of the sample and represent it as a function of irradiation time,
as in the main part of figure \ref{fig:kpcir}. It is easy to see
that the temperature of the sample dropped by around 0.7º C after
25 minutes of laser irradiation. Interestingly, after those 25 minutes
we shut off the laser and this shows up in this figure as an upturn
in the temperature of the sample. It is important to stress that,
even though it could look like a small temperature change, this is
actually quite a good result if one takes into account that there
is only a 0.5\% Er concentration and no attempt was made to optimize
the cooling process.

An interesting fact about these measurements is that there was upconverted
emission present during the realization of all these experiments that
could be observed with the naked eye. In order to investigate whether
this upconversion played a role in the upconversion process, we used
a simple rate equation formalism that has been described elsewhere
\cite{key-31,key-32}. The main conclusion coming out from such an
analysis is that the efficient upconversion present in these materials
provides an additional channel for extracting energy from the system
and, therefore, the cooling efficiency increases. Moreover, when upconversion
is present, the onset of cooling, that is, the frequency below which
cooling occurs, is larger than in the absence of upconversion and,
in some limiting cases, there could be cooling even in the Stokes
region. The interested reader is referred to those references for
more details.

\section{Luminescence control with photonic nanostructures}

In this section I describe what I think it is the next step into improving
the luminescence efficiency of solids and, in particular, the optical
cooling efficiency: the use of RE-doped periodic nanoscale photonic
heterostructures --also known as photonic crystals-- doped with optically
active ions for controlling the density of photonic states.

A photonic crystal is an engineered periodic structure made of two
or more materials with very different dielectric constants. These
photonic nanostructures have generated an ever-increasing interest
in the last twenty years because of their potential to control the
propagation of light to an unprecedented level \cite{key-33,key-34,key-35,key-36,key-37,key-38,key-39,key-40,key-41,key-42,key-43,key-44,key-45,key-46,key-47,key-48,key-49}.
When an electromagnetic wave (EM) propagates in such a structure whose
period is comparable to the wavelength of the wave, interesting phenomena
occur. Among the most interesting ones are the possibility of forming
a complete photonic band gap (CPBG), that is, a frequency range for
which no photons having frequencies within that range can propagate
through the photonic crystal, to localize light by introducing several
types of defects in the lattice, or enhancing certain non-linear phenomena
due to small group velocity effects.%
\begin{figure}
\begin{centering}\includegraphics{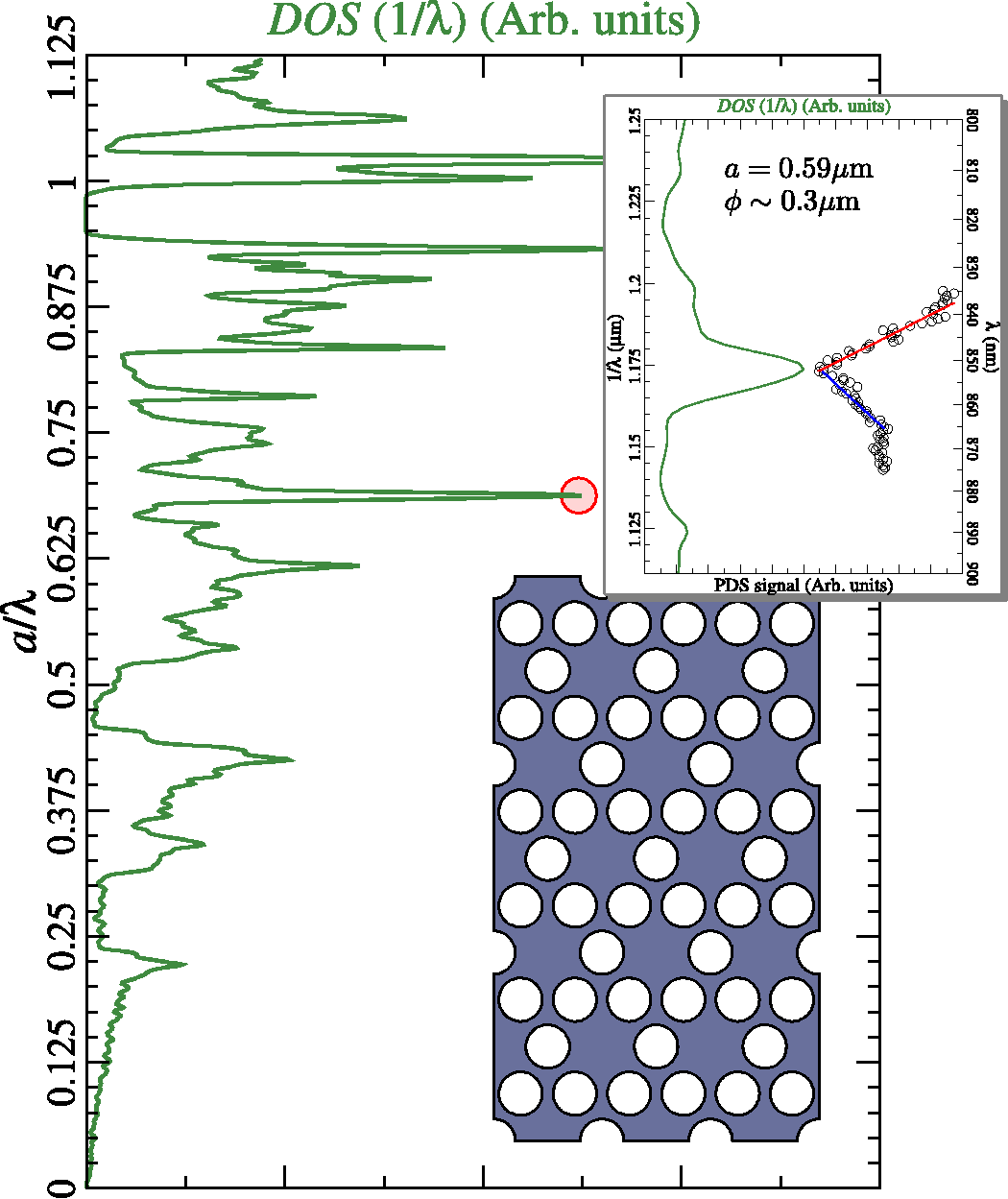}\par\end{centering}
\caption{\label{fig:dos}Photonic density of states of a two-dimensional photonic
crystal based on an inverse kagomé lattice (bottom inset). The upper
inset shows the maximum marked with a small circle for the lattice
parameter $a=590$ nm and diameter of the air holes $\phi=300$ nm
together with the PDS signal for Er-doped KPC crystal.}
\end{figure}

More important for our present purpose, a photonic crystal doped with
an OAI offers new pathways in which one can improve the luminescence
efficiency of the ions embedded in such a structure and, in particular,
the cooling efficiency for laser cooling applications: firstly, they
allow to control the spontaneous emission of those ions. To see this,
one needs to take into account that the probability per unit time
of an spontaneous emission of a photon is given by\begin{equation}
\omega_{sp}(\nu)=\frac{2\pi}{\hbar}\left|M_{ed}\right|^{2}DOS(\nu),\end{equation}
where $M_{ed}$ is the electric dipole photon-ion coupling and $DOS(\nu)$
the density of photonic states in which the ion can decay. The important
point to notice here is the fact that this emission rate is proportional
to the photonic density of states. In a homogeneous medium, this quantity
follows a rather uninteresting power law. However, in a photonic crystal,
this is a highly non-trivial quantity that possesses a lot maxima
and minima \cite{key-54}, such as the one shown in figure \ref{fig:dos}
for a two-dimensional crystal based on an inverse kagomé lattice \cite{key-50,key-51}.
The idea is then to adjust the geometrical parameters of the photonic
crystal in such a way that one of these maxima coincides with the
barycenter of the emitting level of the OAI, so that the power radiated
out from the system as spontaneous emission maximizes and so it does
the cooling efficiency of the system. For example, in the particular
case of the photonic crystal displayed in figure \ref{fig:dos}, choosing
a lattice parameter $a=590$ nm and a diameter for the air holes $\phi=300$
nm, we would match one of the maxima in the photonic DOS with the
barycenter of the $^4$I$_{9/2}$ level than can be easily identified
in the PDS signal, thus maximizing the spontaneous emission rate of
this band.

Secondly, another effect we should take into account is the fact that
the constituents of the photonic crystal (dielectric spheres, for
example) are of finite size, a few hundred nanometer diameter typically
for applications in the visible and near infrared frequency range,
so that this leads to important finite size effects in the \emph{phononic}
density of states. In fact, this has been recently investigated by
Ruan and Kaviany \cite{key-56} in RE-doped nanocrystalline powders
made of small clusters of diameter around 3 nm. These authors found
that the long wavelength region --small energy-- of the phonon density
of states is enhanced when compared when the corresponding bulk crystal
counterpart. This means that we have more phonons of small energy
and, therefore, the probability of having an anti-Stokes absorption
event increases which, in turn, increases the cooling efficiency.
Given the similarity between the systems analyzed by Ruan and Kaviany
and the photonic crystals considered in this work with regards to
the finite size of the constituents, one can expect that a similar
line of reasoning applies to these later structures.

Finally, we could also use point defects introduced in a controlled
way into the photonic crystal to improve the cooling efficiency. It
is well known that light tends to localize around these defects in
the presence of a complete photonic band gap \cite{key-33}. As a
consequence, the effective interaction time between the light and
the OAI increases and so it does the total probability of having an
anti-Stokes absorption event. Actually, Ruan and Kaviany found that
light localization in RE-doped powders (due to a different physical
phenomenon than the one considered here) by itself is responsible
for an enhancement of about a 100\% of the cooling efficiency \cite{key-56}.

Of course, the best route to maximize the enhancement of the cooling
efficiency due to these effects is to take all of them into account
at the same time. That is a venue of research that our group is currently
pursuing.

\section{Conclusions}

As mentioned in the introduction, this manuscript summarizes three
years of research carried out by our group in the field of the optical
spectroscopy of Er-doped low-phonon potassium lead halides. The results
presented here clearly show that these materials are very promising
for ir-to-vis upconversion applications: their small phonons make
upconversion very efficient in these systems, they exhibit a wealth
of upconversion mechanisms so that the dynamics is complex in spite
of their relative simplicity, and the type of dynamics can be easily
selected by adequately tuning the pumping laser frequency.

I have also described how one can use conventional optical processes
(anti-Stokes absorption) for non-conventional applications: laser
cooling of luminescent solids. Indeed, potassium lead chloride is
a very useful host material for this type of applications due to its
extremely small phonons. In the past, we demonstrated laser cooling
in an Yb$^{3+}$-doped KPC crystal \cite{key-29}. More recently,
we demonstrated laser cooling in an Er$^{3+}$-doped KPC crystal \cite{key-26},
with the special significance this has for potential applications
in the field of optical telecommunications and for bioimaging and
phototherapy. This later result is briefly described here together
with a novel upconversion cooling mechanism we have recently identified.

Finally, I have made a discursion into --what I consider-- is the
logical next step in the field of optical spectroscopy of condensed
phases: controlling the luminescence efficiency by using photonic
nanoscale heterostructures as host materials for optically active
ions. I have used the cooling efficiency as a guiding example of this
idea and I have explained the different ways in which this quantity
can be enhanced by using a photonic crystal doped with optically active
ions: control of the spontaneous emission rate of the optically active
ion by means of the photonic density of states of the photonic crystal,
enhancement of the phononic density of states in the long wavelength
region due to the finite size of the constituents of the photonic
crystal, and introduction of point defects in a controlled way to
localize light around them.

I hope these pages serve the purpose of motivating other researchers
to further explore these ideas so we can advance into this new era
of the optical spectroscopy in which not only will we be able to understand
the optical properties of the condensed phases but also to tailor
them for specific applications.

\begin{acknowledgments}
I got acquainted with Professor Michael Sturge's works when, as a
graduate student, I was working on elucidating the mechanisms responsible
for the anomalous low-temperature behavior of the homogeneous linewidths
of amorphous solids doped with optically active ions. I admired the
well reputed scientist that, among other contributions, settled the
foundations that allowed us to understand this phenomenon. It is thus
easy to understand the great and unexpected honor it has been for
me to receive this prize named after him. Obviously, there have been
many people that helped me during these long years: Joaquín Fernández,
Rolindes Balda, and all my colleagues at the Dpto.~de Fisica Aplicada
I, Miguel Angel Cazalilla, Dave Huber, Steve Conradson, and Bill Yen
--a very special friend. But, above all of them, I would like to dedicate
this prize to my wife, who has given me love and support during all
these years.

Of course, the results presented in this manuscript are not the work
of a single person. Rather, they are the outcome of a strong collaboration
between Joaquín Fernández, Rolindes Balda, and myself at the Dpto.~de
Física Aplicada I of the E.T.S.~de Ingeniería de Bilbao. Joaquín
and Rolindes have carried out the hard experimental work described
above and to them goes my gratitude for this and for the many things
I have learned from them through the years.

Finally, I acknowledge financial support from the Spanish Ministerio
de Educación y Ciencia under the Ramón y Cajal program.
\end{acknowledgments}

\end{document}